\documentclass[letterpaper, 10 pt, conference]{ieeeconf}

\IEEEoverridecommandlockouts
\usepackage{amsmath,amssymb,amsfonts}
\usepackage{algorithmic}
\usepackage{graphicx}
\usepackage{textcomp}
\usepackage{xcolor}
\usepackage{amsmath,amssymb,amsfonts}
\usepackage{algorithmic}
\usepackage{graphicx}
\usepackage{textcomp}
\usepackage{xcolor}
\newtheorem{definition}{Definition}
\newtheorem{theorem}{Theorem}
\newtheorem{remark}{Remark}
\newtheorem{problem}{Problem}

\usepackage{units}

\usepackage{booktabs}

\DeclareMathOperator{\rank}{rank}

\def\BibTeX{{\rm B\kern-.05em{\sc i\kern-.025em b}\kern-.08em
    T\kern-.1667em\lower.7ex\hbox{E}\kern-.125emX}}

\definecolor{vgRed}{RGB}{193, 48, 24}
\definecolor{vgOrange}{RGB}{243, 111, 19}
\definecolor{vgYellow}{RGB}{235, 203, 56}
\definecolor{vgGreen}{RGB}{162, 185, 105}
\definecolor{vgLightBlue}{RGB}{13, 149, 188}
\definecolor{vgDarkBlue}{RGB}{6, 56, 81}

\usepackage{tikz}
\usetikzlibrary{decorations.markings}

\usepackage{pgfplots}
\pgfplotsset{compat=1.13}
\usepgfplotslibrary{statistics}

\usepgfplotslibrary{external}
\tikzset{external/system call={pdflatex \tikzexternalcheckshellescape -halt-on-error
        -interaction=batchmode -jobname "\image" "\texsource"}}
\pgfplotsset{every axis/.append style={semithick,tick style={major tick
      length=4pt,semithick,black}}}

\pgfkeys{/pgfplots/x axis shift down/.style={
    x axis line style={yshift=-#1},
    xtick style={yshift=-#1},
    tick align=outside,
    xticklabel shift={#1}}}

\pgfkeys{/pgfplots/y axis shift left/.style={
    y axis line style={xshift=-#1},
    ytick style={xshift=-#1},
    yticklabel shift={#1},
    scaled y ticks = false,
      tick align=outside,
    y tick label style={/pgf/number format/fixed,
    }
  }
}

\pgfplotsset{linestyle boxplot blue/.style={%
  boxplot,
  boxplot/box extend=0.4,
  mark=*,
  every mark/.append style={mark size=0.7pt, line width=0pt, opacity=0.2, fill=vgLightBlue}, draw=vgLightBlue,
  }
}

\pgfplotsset{linestyle boxplot red/.style={%
  boxplot,
  boxplot/box extend=0.4,
  mark=*,
  every mark/.append style={mark size=0.7pt, line width=0pt, opacity=0.2, fill=vgRed}, draw=vgRed,
  }
}
\pgfplotsset{boxplot legend/.style={
    legend image code/.code={
    \begin{scope}[scale=0.35]
        \draw[#1] (0cm,0cm) rectangle (0.6cm,0.3cm)
        (0.3cm,0cm) -- (0.3cm,-0.1cm) (0.1cm,-0.1cm) -- (0.5cm,-0.1cm)
        (0.3cm,0.3cm) -- (0.3cm,0.4cm) (0.1cm,0.4cm) -- (0.5cm,0.4cm);
    \end{scope}
    },
}}

\pgfplotsset{boxplot semilogy/.style={
    height = 5.46cm,
    width = 8.5cm,
    line width = 0.7pt,
    separate axis lines,
    axis x line*=bottom,
    x axis shift down = 3pt,
    enlarge x limits=false,
    axis y line*=left,
    y axis shift left = 15pt,
    enlarge y limits={abs=.25pt},
    enlarge x limits={abs=.25pt},
    boxplot/draw direction = y,
    ymin = 1e-12,
    ytick = {1e-18, 1e-15, 1e-12, 1e-9, 1e-6, 1e-3, 1e0},
    x axis line style={white},
    xtick style={white},
    xticklabel style={yshift = 2mm},
    xtick={1.5, 3.5, ..., 21.5},
    xticklabels={1, 2, ..., 10},
    clip=true,
    ylabel = {Prediction error},
    y label style={at={(-0.08, 1.05)}, anchor=north east},
    xlabel = {Prediction step},
    legend columns=2,
    legend style={
      at={(0.5, 1.0)},
      anchor=south,
      draw=none,
      fill=none,
      legend cell align=left,
      /tikz/every even column/.append style={column sep=0.3cm}
      },
}}

\usepackage{todonotes}
\makeatletter
\renewcommand{\todo}[2][]{\tikzexternaldisable\@todo[#1]{#2}\tikzexternalenable}
\newcommand{\atJl}[1]{\tikzexternaldisable\@todo[inline, color=vgLightBlue]{@Johannes: #1}\tikzexternalenable}
\newcommand{\atCh}[1]{\tikzexternaldisable\@todo[inline, color=vgYellow]{@Christian: #1}\tikzexternalenable}
\newcommand{\atAll}[1]{\tikzexternaldisable\@todo[inline, color=vgGreen]{@All: #1}\tikzexternalenable}
\renewcommand{\missingfigure}[2][]{\tikzexternaldisable\@missingfigure[#1]{#2}\tikzexternalenable}
\makeatother

\usepackage{acronym}

\acrodef{pv}[PV]{photovoltaic}
\acrodef{mg}[MG]{microgrid}
\acrodef{mpc}[MPC]{model predictive control}
\acrodef{lti}[LTI]{linear time-invariant}
\acrodef{dd}[DD]{data-driven}
\acrodef{res}[RES]{renewable energy sources}
\acrodef{i/o}[I/O]{input/output}
\acrodef{bsu}[BSUs]{Battery storage units}

\IEEEoverridecommandlockouts                              

\title{\LARGE \bf Data-driven model predictive control of \mbox{battery storage units}
}

\author{Johannes B. Lipka$^{1}$ and Christian A. Hans$^{2}$
\thanks{$^{1}$Johannes B. Lipka is with Siemens AG, Germany, {\tt\small johannes-bernd.lipka@siemens.com}}%
\thanks{$^{2}$Christian A. Hans is with the University of Kassel, Germany, {\tt\small hans@uni-kassel.de}}%
}

\begin{document}

\maketitle

\begin{abstract}
    In many state-of-the-art control approaches for power systems with storage units, an explicit model of the storage dynamics is required.
    With growing numbers of storage units, identifying these dynamics can be cumbersome. 
    This paper employs recent data-driven control approaches that do not require an explicit identification step.
    Instead, they use measured input/output data in control formulations.
    In detail, we propose an economic data-driven \ac{mpc} scheme to operate a small power system with input-nonlinear battery dynamics.
    First, a linear data-driven \ac{mpc} approach that uses a slack variable to account for plant-model-mismatch is proposed.
    In a second step, an input-nonlinear data-driven \ac{mpc} scheme is deduced.
    Comparisons with a reference indicate that the linear data-driven \ac{mpc} approximates the nonlinear plant in an acceptable manner.
    Even better results, however, can be obtained with the input-nonlinear data-driven \ac{mpc} scheme which provides increased prediction accuracy.
\end{abstract}

\acresetall


\section{Introduction}
Battery storage units are integral to future power systems.
They exhibit diverse dynamics based on technology and production. 
Moreover, their dynamics may change over time due to ageing.
Effective control of future power systems with storage units necessitates consideration of these dynamics, typically through a system identification step before control synthesis. 
In systems with multiple batteries, each with different dynamics, identifying and updating their models can be cumbersome. 
This motivates alternative data-driven control schemes which do not require an explicit identification step.

Willems et al. \cite{WillemsTheorem} provided foundations for modeling of \ac{lti} systems using \ac{i/o} data.
In the context of behavioural systems, they proved that if the input signal of an \ac{lti} system is persistently exciting (see Section \ref{sec: Mathematical Preliminaries}), then the Hankel matrix of sufficiently many \ac{i/o} measurements can span the vector space of all possible trajectories of the system.
The authors of \cite{Algoewer_StatSpaceOfWillem,WillemToStateSpace2,MPC_FeedbackLinearizable,Controller_FeedbackLinearizable} extended the results of \cite{WillemsTheorem} to \ac{lti} state models and to feedback-linearizable Hammerstein and Wiener systems.
In \cite{LygerosGridConnectedConverterDDControl,Lygeros_InTheShallowsOfDeePC,Lygeros_RegularizedRobustDDControl} the theory was further extended to unknown \ac{lti} systems, ensuring trajectory tracking and constraint satisfaction under some easy-to-hold assumptions.
The authors of \cite{DDControlWithStabilityGuarantees} presented a trajectory-based \ac{mpc} scheme with terminal constraints and could find conditions for exponential stability in presence of measurement noise.
In \cite{Linear_Tracking_MPC_1,Linear_Tracking_MPC_2}, these conditions were extended for slow changing nonlinear systems. 

Despite recent advancements in data-driven control theory and their application to practical settings, the utilization of data-driven approaches in economic \ac{mpc} schemes has just recently started (see \cite{DD_multi_energy_dis_system,Xie2022,HVAC_Example}).
To the authors' best knowledge, data-driven \ac{mpc} also has not been applied to plants with battery storage systems. 
Furthermore, there exists only a small number of data-driven control approaches for power systems \cite{LygerosGridConnectedConverterDDControl,DD_multi_energy_dis_system,MAHDAVIPOUR202291, dd_descriptor}.
Moreover, the extended fundamental lemma from \cite{Algoewer_StatSpaceOfWillem} has been rarely employed in an \ac{mpc} context up to know.
 
In this paper, we aim to further bridge the gap between data-driven control theory and applications in the power systems domain. 
Our contributions include:
\begin{itemize}
    \item Developing different economic data-driven \ac{mpc} approaches to control an islanded power system with nonlinear Hammerstein-type battery dynamics.
    In detail, we propose one  data-driven \ac{mpc} approach that considers \ac{lti} dynamics and one which considers input-nonlinearities with a known structure. 
    \item Showcasing how the \textit{extended} fundamental lemma of \cite{Algoewer_StatSpaceOfWillem} can be practically employed in an \ac{mpc} formulation to control battery storage systems. 
    \item Quantitatively comparing the prediction capabilities of the different \ac{mpc} approaches with a reference controller that has perfect knowledge of the system's nonlinearities in case studies. More specifically, we evaluate the prediction accuracy and associated constraint violations of the proposed \ac{mpc} schemes.
\end{itemize}

The paper is structured as follows. 
Section \ref{sec: Mathematical Preliminaries} focuses on mathematical preliminaries.
In Section \ref{sec: Microgrid Model}, the running example of an islanded grid with input-nonlinear battery dynamics is introduced. 
The cost function and a reference \ac{mpc} are discussed in Section \ref{sec: reference mpc}.
A linear data-driven \ac{mpc} scheme is presented in Section \ref{sec: linear mpc}.
In Section \ref{sec: Hammerstein type data-driven MPC}, a nonlinear data-driven \ac{mpc} scheme that deals with Hammerstein-type system dynamics of known structure is presented.
Section \ref{sec: conclusions} concludes the work.


\section{Mathematical Preliminaries} \label{sec: Mathematical Preliminaries}
In what follows, first the notation is discussed.
Then, Hankel matrices and persistence of excitation are defined and Willems' fundamental lemma for \ac{lti} systems is presented. 
Finally, Hammerstein systems are introduced and the extension of Willems' fundamental lemma for feedback-linearizable systems is recalled. 
\subsection{Notation}
Let the set of real numbers, nonnegative real numbers, positive real numbers and negative real numbers be $\mathbb{R}$, $\mathbb{R}_{\geq0}$, $\mathbb{R}_{>0}$ and $\mathbb{R}_{<0}$, respectively.
Integers, nonnegative integers and positive integers are denoted by $\mathbb{Z}$, $\mathbb{N}_{0}$ and $\mathbb{N}$, respectively.
Let $\mathbb{X}=\{x(k)\}_{k=k_{A}}^{k_{B}}$ denote a sequence, i.e., an enumerated and ordered set $\{x(k_{A}),\dots,x(k_{B})\}$ of elements $x(k)\in\mathbb{R}^{n}$ with $k_{A}\in\mathbb{N}_{0}$, $k_{B}\in\mathbb{N}_{0}$ and $k_{A}\leq k_{B}$.
The notation $x(k|t)$ is used to refer to a prediction performed at time instant $t\in \mathbb{N}_{0}$ for prediction step $k\in \mathbb{N}_{0}$, i.e., time instant $t+k$.
$X=[x(k)]_{k=k_{A}}^{k_{B}}$ is shorthand for $[x(k_{A})^{T}~\cdots~x(k_{B})^{T}]^{T}$.
Moreover, $|a|$ refers to the absolute value of $a\in\mathbb{R}$. 
\subsection{Hankel Matrix}
A sequence $\mathbb{X}=\{x(k)\}_{k=0}^{N-1}$ of elements $x(k)\in \mathbb{R}^n$ of length $N\in\mathbb{N}$ can induce a Hankel matrix of order $L\in\mathbb{N}$, $L<N$, of the form
\begin{equation}
    H_{L}(\mathbb{X}) := \begin{bmatrix}
x(0) & x(1)& \cdots & x(N-L)\\
x(1) & x(2)& \cdots & x(N-L+1)\\
\vdots & \vdots & \ddots & \vdots\\
x(L-1) & x(L)& \cdots & x(N-1)
\end{bmatrix}.
\end{equation}
In detail, $H_{L}(\mathbb{X})$ is constructed by cutting the sequence $\mathbb{X}$ into $N-L+1$ snippets of length $L$.
Each column of $H_{L}(\mathbb{X})$ represents a snippet which is shifted one time step ahead compared to the column left of it.
To proceed with what follows, we recall \cite[Definition 1]{Algoewer_StatSpaceOfWillem}.
\begin{definition}[Persistence of Excitation]\label{def: persistence of Excitation}
\textit{A sequence $\mathbb{X}$=$\{x(k)\}_{k=0}^{N-1}$ with $x(k) \in \mathbb{R}^{n}$ is persistently exciting of order $L$ if}
\begin{equation}
    \rank(H_{L}(\mathbb{X})) = nL \hspace{0.1cm} .
\end{equation}
\end{definition}
\begin{remark} \label{remark: L tilde}
    If a sequence is persistently exciting of order $L$, then it is also persistently exciting of order $\tilde{L}\leq L$ \cite{DDControlWithStabilityGuarantees}.
\end{remark}

\subsection{Data-Driven Representation of \acs{lti} Systems}
Let $G$ be a controllable \ac{lti} system and
\begin{subequations}\label{eq:ltiStateSpace}
    \begin{align}
        x(k+1) &= Ax(k) + Bu(k),\\
        y(k) &= Cx(k) + Du(k),
    \end{align}
\end{subequations}
with $x(0)=x_{0}$ be a minimal realization of $G$. 
Here, $x(k)\in\mathbb{R}^{n}$ is the state vector,  $y(k)\in \mathbb{R}^{p}$ the output vector and $u(k)\in \mathbb{R}^{d}$ the input vector.
Measuring the discrete-time input and output signals of (\ref{eq:ltiStateSpace}) for $N$ consecutive times allows us to form the input and output sequences $\mathbb{U}^{m}=\{u^{m}(k)\}^{N-1}_{k=0}$ and $\mathbb{Y}^{m}=\{y^{m}(k)\}^{N-1}_{k=0}$.  
If $\mathbb{U}^{m}$ is persistently exciting of order $L$, then the following theorem from \cite[Theorem 3]{Algoewer_StatSpaceOfWillem} which is based on \cite{WillemsTheorem} holds.
\begin{theorem} \label{theorem: Linear Case}
\textit{An input sequence} $\{u(k)\}^{L-1}_{k=0}$ \textit{and its corresponding output sequence} $\{y(k)\}^{L-1}_{k=0}$ \textit{represent a trajectory of $G$ if and only if there exists a vector} $\alpha\in \mathbb{R}^{N-L+1}$ \textit{such that}
\textit{with} $\mathbf{U}=[u(k)]^{L-1}_{k=0}$ \textit{and} $\mathbf{Y}=[y(k)]^{L-1}_{k=0}$
\textit{it holds that}
\begin{equation} \label{eq: DD Trajectory Criteria}
\begin{bmatrix}
\mathbf{U}\\
\mathbf{Y}
\end{bmatrix}=\begin{bmatrix}H_{L}(\mathbb{U}^{m})\\
H_{L}(\mathbb{Y}^{m})
\end{bmatrix}\alpha.
\end{equation}
\end{theorem}
In other words, the space of all possible \ac{i/o} trajectories of $G$ is spanned by the matrix $\begin{bmatrix}
H_{L}(\mathbb{U}^{m})^{T} \hspace{0.3cm}
H_{L}(\mathbb{Y}^{m})^{T}
\end{bmatrix}^{T}$.
Thus, the \ac{i/o} behavior of $G$ can be described by time-shifted windows of measured \ac{i/o} signals supposed that $\mathbb{U}^{m}$ is persistently exciting. 
Note that the persistence of excitation criterion does not need to be checked for the output sequence $\mathbb{Y}^{m}$ \cite{WillemsTheorem,Algoewer_StatSpaceOfWillem}.

When forecasting future trajectories of a system, the initial conditions need to be included.
In a minimal realization of an \ac{lti} system $G$ of order $n\in\mathbb{N}$, the associated input and output sequences $\mathbb{U}$ and $\mathbb{Y}$ of length larger than or equal to $n$ induce a unique internal state trajectory \cite{Algoewer_StatSpaceOfWillem}.
This property can be used to describe the initial state:
Let $\tilde{n}\ge n$, $\tilde{n}\in\mathbb{N}$ represent an upper approximation of the order of $G$. 
Then, an input sequence $\mathbb{U}^{m}$ and its corresponding output sequence $\mathbb{Y}^{m}$ of length $\tilde{n}$ (or larger) describe a unique initial state of~$G$.

For example, the initial conditions $x(t)=x_{t}$ of a  minimal realization of $G$ can be described by the $\tilde{n}$ most recent \ac{i/o} measurements at time $t\ge \tilde{n}$.
That is, the sequences $\mathbb{U}_{t}^{m} = \{u^{m}(k|t)\}^{-1}_{k=-\tilde{n}}$ and $\mathbb{Y}_{t}^{m} = \{y^{m}(k|t)\}^{-1}_{k=-\tilde{n}}$ induce the unique initial state $x(t)$.
In conclusion, we can adequately describe the behavior and initial state of an \ac{lti} system using \ac{i/o} data and an upper approximation of its order $n$~\cite{Algoewer_StatSpaceOfWillem}.
\subsection{Data-Driven Representation of Hammerstein Systems} \label{sec: hammerstein}
Consider a Hammerstein system of the form
\begin{subequations}\label{eq:hammersteinSystem}
    \begin{align}
        x(k+1) &= Ax(k) + B\psi(u(k)), \label{eq: Hammerstein System1}\\
        y(k) &= Cx(k) + D\psi(u(k)), \label{eq: Hammerstein System2}
    \end{align}
\end{subequations}
with $x(0) = x_{0}$ where $\psi: \mathbb{R}^{d}\rightarrow \mathbb{R}^{\tilde{d}}$ is a nonlinear mapping of input $u$ from $\mathbb{R}^{d}$ to $\mathbb{R}^{\tilde{d}}$ .
If $\tilde{d}=1$, then $\psi$ takes the form
\begin{equation} \label{eq: psi}
    \psi(u) = \alpha_{1}\psi_{1}(u)+\alpha_{2}\psi_{2}(u)+\ldots+\alpha_{r}\psi_{r}(u)
\end{equation}
where $\psi_{i}:\mathbb{R}^{d}\rightarrow\mathbb{R}$ are linear and nonlinear mappings and $\alpha_{i}\in\mathbb{R}$ coefficients which are nonzero for at least one $i\in\{1,\dots,r\}$ \cite{Algoewer_StatSpaceOfWillem}.
Let $G$ be a Hammerstein system of the form (\ref{eq:hammersteinSystem}). 
For simplicity, let $\tilde{d}=1$ such that $\psi$ can be described by (\ref{eq: psi}).
Then, a new auxiliary linear input vector
\begin{equation} \label{eq: linear-input}
    v(k) = \begin{bmatrix}
    \psi_{1}(u(k)) &
    \psi_{2}(u(k)) &
    \cdots &
    \psi_{r}(u(k))
    \end{bmatrix}^{T}
\end{equation}
can be formed. 
This vector can then be used in following theorem from \cite[Proposition 5]{Algoewer_StatSpaceOfWillem}.

\begin{theorem} \label{theorem: Nonlinear Case}
\textit{Suppose that} $\mathbb{U}^{m}=\{u^{m}(k)\}^{N-1}_{k=0}$ \textit{and} $\mathbb{Y}^{m}=\{y^{m}(k)\}^{N-1}_{k=0}$ \textit{are measured \ac{i/o} sequences of the Hammerstein system }(\ref{eq:hammersteinSystem}) \textit{with order} $n$. \textit{ Moreover, let} $\mathbb{V}^{m}=\{v^{m}(k)\}^{N-1}_{k=0}$ \textit{be the corresponding linear input sequence with $v(k) = [
    \psi_{1}(u^{m}(k))^{T}~
    \psi_{2}(u^{m}(k))^{T}~
    \cdots ~
    \psi_{r}(u^{m}(k))^{T}]^{T}$}.
\textit{If} $\mathbb{V}^{m}$ \textit{is persistently exciting of order} $L$, \textit{then an \ac{i/o} sequence } $\{v(k)\}^{L-1}_{k=0}$, $\{y(k)\}^{L-1}_{k=0}$ \textit{is a trajectory of system \eqref{eq:hammersteinSystem} if and only if there exists a vector} $\alpha\in \mathbb{R}^{N-L+1}$ \textit{such that}
\begin{equation} \label{eq: DD Trajectory Criteria Hammerstein}
\begin{bmatrix}
\mathbf{V}\\
\mathbf{Y}
\end{bmatrix}=\begin{bmatrix}H_{L}(\mathbb{V}^{m})\\
H_{L}(\mathbb{Y}^{m})
\end{bmatrix}\alpha
\end{equation}
\textit{where}
$\mathbf{Y}=[y(k)]^{L-1}_{k=0}$
\textit{and} 
$\mathbf{V}=[v(k)]^{L-1}_{k=0}$ \textit{with} $v(k)$ \textit{from} \eqref{eq: linear-input}.
\end{theorem}


\section{Model of an islanded grid} \label{sec: Microgrid Model}

\begin{figure}[h]
    \centering
    \tikzsetnextfilename{testSetupWithLoad}

\newcommand{\lnwidth}{0.6pt}
\newcommand{\lngrid}{0.100pt}

\newcommand{\scaleSymbols}{0.03}
\input{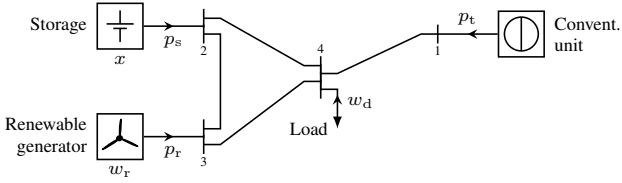}

\newcommand{\wInPlot}[1]{\ensuremath{w_{\mathrm{#1}}}}
\newcommand{\wDPlot}{\wInPlot{d}}
\newcommand{\wRPlot}{\wInPlot{r}}

\newcommand{\xSPlot}{\ensuremath{x}}
\newcommand{\pInPlot}[1]{\ensuremath{p_{\mathrm{#1}}}}
\newcommand{\pTPlot}{\pInPlot{t}}
\newcommand{\pSPlot}{\pInPlot{s}}
\newcommand{\pRPlot}{\pInPlot{r}}
\newcommand{\pEPlot}[1]{\ensuremath{p_{\mathrm{e},#1}}}

\newcommand{\uInPlot}[1]{\ensuremath{u_{\mathrm{#1}}}}
\newcommand{\uTPlot}{\uInPlot{t}}
\newcommand{\uSPlot}{\uInPlot{s}}
\newcommand{\uRPlot}{\uInPlot{r}}

\tikzstyle{myStyle} = [scale=.7, >=latex,
	rotate=90,
	line width=\lnwidth, 
	font=\scriptsize,
	invS/.style={rectangle, fill=white, draw=black, inner sep=0.1pt, line width=\lnwidth} ]

\tikzset{midArrow/.style={
		line width=\lnwidth,
		draw,%
		decoration={%
			markings,%
			mark=at position 0.5 with \arrow{stealth},%
		},%
		postaction=decorate}}%

\begin{tikzpicture}[myStyle, color=black, xscale=.6, yscale=1.27]

\path (-2+0.25, 3.0) node[invS] (nodeR) {\GraphTristateR}
	node[at = (nodeR.180), anchor=east, align=right] {Renewable \\ generator};
\path[midArrow] (nodeR) -- node[below]{\pRPlot} (-2+0.25, 1.75);

\path (2-0.25, 3.0) node[invS] (nodeS) {\GraphTristateS}
	node[at = (nodeS.180), anchor=east] {Storage};
\path[midArrow] (nodeS) -- node[below]{\pSPlot} (2-0.25, 1.75);

\path[draw, line width=\lnwidth] (-2, 1.75) node{\GraphLineFlip}  node[below, font=\tiny, yshift=-1]{3} --
	(-2, 1.5) --
	(0.0, 0.25) --
	(0.0, 0) node{\GraphLineFlip};

\path[draw, line width=\lnwidth] (2, 1.75) node{\GraphLineFlip} --
	(2, 1.5) --
	(0.5, 0.25) --
	(0.5, 0) node{\GraphLineFlip} node[above, font=\tiny, yshift=1]{4};

\path[draw, line width=\lnwidth] (1.5, 1.75) node{\GraphLineFlip} node[below, font=\tiny, yshift=-1]{2} --
	(1.5, 1.5) --
	(-1.5, 1.5) --
	(-1.5, 1.75) node{\GraphLineFlip} ;

\path (1.5, -3) node[invS] (nodeT) {\GraphTristateT}
	node[at = (nodeT.0), anchor=west, align=left] {Convent. \\ unit};

\path[midArrow] (nodeT) -- node[above]{\pTPlot} (1.5, -1.75);
\path[draw, line width=\lnwidth] (1.5, -1.75) node{\GraphLineFlip} node[below, font=\tiny, yshift=-1]{1} --
(1.5, -1.5) --
(0.25, -0.25) -- (0.25, 0) node{\GraphLineFlip};

\path[draw, line width=\lnwidth] (-0.25, 0) node{\GraphLineFlip};
\path[<-,
	draw, line width = \lnwidth,
	decoration={%
		markings,%
		mark=at position 0.6 with \arrow{stealth},%
	},%
	postaction=decorate]
	(-1.5, -0.25) node[left]{Load} -- (-0.25, -0.25) node[right, yshift=-5]{\wDPlot} -- (-0.25, 0.0);

\node[below = -0.5mm of nodeR, yshift=-0] {\wRPlot};
\node[below = -0.5mm of nodeS, yshift=-0] {$x$};

\end{tikzpicture}%
     \caption{Islanded grid  composed of storage unit, conventional generator, renewable generator and a load. Layout from \cite{christian_hans}.}
    \label{fig: Simple Microgrid with PL Example}
\end{figure}

The following section presents a minimal example of an islanded grid which is considered throughout this work to illustrate how data-driven models of battery dynamics can be employed.
Note that the presented results are not restricted to this structure and can be easily extended to more complex ones. 
Figure~\ref{fig: Simple Microgrid with PL Example} depicts the grid consisting of a load, an energy storage unit, a conventional and renewable generator with power $w_{d}$, $p_{s}$, $p_{t}$, $p_{r}$, respectively.
Arrows indicate power direction: positive values describe power provided to the grid, negative values describe power consumed. 

\subsection{Renewable Generator}

The power of the renewable generator can be modeled by
\begin{equation} \label{eq: wind turbine power output}
    p^{min}_{r} \leq p_{r}(k) \leq w_{r}(k),
\end{equation}
where $w_{r}(k)\in \mathbb{R}$ is the weather-dependent available renewable power at time-step $k\in\mathbb{N}_{0}$ and $p^{min}_{r}\in\mathbb{R}_{\ge0}$ the minimum power.
Note that in \eqref{eq: wind turbine power output} the renewable infeed $p_{r}(k)$ is assumed to be curtailable to values below the available power $w_{r}(k)$. 

\subsection{Conventional Unit}
We consider a conventional unit whose change in output power $\Delta p_{t}(k) = p_{t}(k)-p_{t}(k-1)$ is not restricted. 
The on/off status of the unit is represented by a binary variable $\delta(k)$, i.e.,
\begin{equation*}
\delta(k)=\begin{cases}
			0, & \text{if the conventional unit is disabled},\\
            1, & \text{if the conventional unit is enabled}.
		 \end{cases}
\end{equation*}
If $\delta(k)=0$, then $p_{t}(k)$ is zero.
Otherwise, $p_{t}(k)$ is bounded by the minimum power  $p^{min}_{t}$$\in$$\mathbb{R}_{\ge0}$ and the maximum power $p^{max}_{t}$$\in$$\mathbb{R}_{>0}$. This can be considered using the constraint
\begin{equation} \label{eq: conventional unit inequality constraint}
p^{min}_{t}\delta(k) \leq p_{t}(k) \leq p^{max}_{t}\delta(k). 
\end{equation}

\subsection{Energy Storage Unit}
Given the availability of economical battery storage units, we focus our modeling efforts on these types of storages.
In detail, we consider storage units whose dynamics can be modeled along the lines of \cite{christian_hans,Simple_Battery_dynamics_1,Simple_Battery_dynamics_2}.
Storage units have physical limitations: the power output is constrained to 
\begin{align}\label{eq: Storage Power Constraints}
   p^{min}_{s} \leq p_{s}(k) \leq p^{max}_{s}
\end{align}
with $p^{min}_{s}\in \mathbb{R}_{<0}$  and  $p^{max}_{s}\in \mathbb{R}_{>0}$.
Recall that if $p_{s}(k)<0$, the battery charges and if $p_{s}(k)>0$, it discharges.
Let $x(k)$ denote the stored energy, which is bounded by 
\begin{equation}\label{eq: Storage Capacity Constraints}
    x^{min} \leq x(k) \leq x^{max}
\end{equation}
with $x^{min}\in \mathbb{R}_{\geq 0}$ and $x^{max}\in \mathbb{R}_{> 0}$. 
We consider nonlinear storage dynamics.
In detail, we use a quadratic loss-term $p_{s}(k)^{2}$ in the dynamics
\begin{subequations}\label{func: Storage Dynamics Quadratic Default}
\begin{align}
x(k+1) &= Ax(k) + B_{l}p_{s}(k) + B_{q}p_{s}(k)^{2}, \\
y(k) &= x(k),
\end{align}
\end{subequations}
to capture conversion losses (AC to DC and vice versa) when charging or discharging \cite{Linearized_Battery_Dynamics_1_Pisano, Linearized_Battery_Dynamics_2_Gholami}.
Matrix $A\in\mathbb{R}_{>0}$ is typically used to model the self discharge.
In realistic settings we have that $0\ll A<1$. 
Moreover, the input matrices $B_{l}\in \mathbb{R}_{<0}$ and $B_{q}\in \mathbb{R}_{<0}$ are negative. 
Note that the dynamics are state-affine but not input-affine.

\subsection{Power Equilibrium}

In an islanded grid, the provided power must equal the consumed power. This can be modeled via
\begin{equation} \label{eq: power equilibrium constraints}
    p_{t}(k) + p_{s}(k) + p_{r}(k) + w_{d}(k) = 0.
\end{equation} 

\section{Reference MPC} \label{sec: reference mpc}
In order to compare and evaluate our novel data-driven \ac{mpc} approaches, we employ a reference \ac{mpc} that has perfect knowledge of the battery dynamics. 
The structure of the reference \ac{mpc} is inspired by \cite{christian_hans}.

In what follows, we will first define a cost function. 
This function is then used to formulate an \ac{mpc} problem which is employed in closed-loop simulations. 
\subsection{Cost Function}
At time-step $t$ and prediction instant $k$, a stage cost
\begin{equation} \label{eq: simple cost function}
    \ell(k|t) = c_{0}(p_{t}(k|t)-p_{r}(k|t)) +c_{1}|\Delta\delta(k|t)| +c_{2}\delta(k|t)
\end{equation}
with weights $c_{1},c_{2},c_{3}\in\mathbb{R}_{>0}$ is considered.
The term $p_{t}(k|t)-p_{r}(k|t)$ rewards renewable infeed and penalizes the use of the conventional unit.
The term $c_{1}|\Delta\delta(k|t)|$ with $|\Delta\delta(k|t)|=|\delta(k|t)-\delta(k-1|t)|$ penalizes the on and off switching of the conventional unit.
Finally, the conventional unit comes with the power-independent cost $c_{2}\delta(k|t)$.

\subsection{Problem Statement}
Let $X=[x(k|t)]^{L}_{k=0}$ and $P=[p(k|t)]^{L-1}_{k=0}$ with $p(k|t)=\begin{bmatrix}
p_{t}(k|t)~p_{s}(k|t)~p_{r}(k|t)  
\end{bmatrix}^{T}$, where $L\in \mathbb{N}$ is the prediction horizon.  
We can then formulate the following optimization problem with certain inputs $w_{r}(k)$ and $w_{d}(k)$.
\begin{problem}[Reference \acs{mpc}] \label{prob:reference}
  \begin{subequations}
    \[
      \min_{P, X} \textstyle\sum\limits_{k=0}^{L-1} \ell(k|t)\cdot\gamma^{k}
    \]
    subject to \eqref{eq: wind turbine power output}--\eqref{eq: Storage Power Constraints}, \eqref{eq: power equilibrium constraints} for all $k\in[0,L-1]$, \eqref{eq: Storage Capacity Constraints} for all $k\in[1,L]$ as well as the dynamics
      \begin{equation} \label{eq: reference l0}
        x(k+1|t) = Ax(k|t) + B_{l}p_{s}(k|t) + B_{q}p_{s}(k|t)^{2}
      \end{equation}
      for all $k\in[0,L-1]$ with the initial conditions
      \begin{equation} \label{eq: reference x0}
        x(0|t) = x^{m}(t),
      \end{equation}
      \begin{equation} \label{eq: reference d0}
      \delta(-1|t) = \delta^{m}(t).
      \end{equation}
  \end{subequations}
\end{problem}

Here, $\delta(-1|t)$ is the current on/off condition of the conventional unit.
Multiplying the predicted cost at time-step $k$ with $\gamma^{k}$, $\gamma\in(0,1)$, allows to put less emphasis on decisions in the far future.
This can be helpful in receding horizon control. 
Note that Problem \ref{prob:reference} is a nonconvex, mixed-integer problem with quadratic constraints in the form of the previously defined dynamics \eqref{eq: reference l0} which can be solved by commercial software such as Gurobi \cite{gurobi}.

At each time step $t$, Problem \ref{prob:reference} is solved. 
From the optimal input trajectory $[p^{*}(k|t)]^{L-1}_{k=0}$, the first value ${p^{*}(0|t)}$ is applied to the system and the system is operated for one time step. Then, measurements for (\ref{eq: reference x0}) and (\ref{eq: reference d0}) are updated and Problem \ref{prob:reference} is solved again in a receding horizon manner.

\subsection{Case Study} \label{sec: reference case study}
\begin{figure}[ht]
    \centering
    \tikzsetnextfilename{linearBatteryResults}
    \input{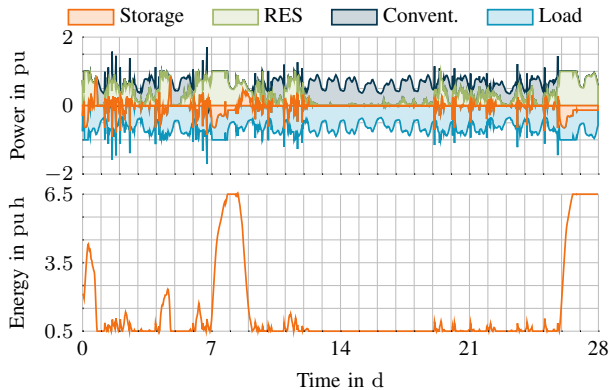}%
    \caption{System trajectory
         with reference \acs{mpc}. (\unit{pu}: per unit, \unit{d}: days)}
    \label{fig: NonlinBat Ref trajectory}
\end{figure}

Consider the grid in Figure \ref{fig: Simple Microgrid with PL Example} with the parameters in Table~\ref{table: Parameter values of simple mpc}.  
Closed-loop simulations over four weeks with a time resolution of 30 minutes (1334 total time steps) were performed using the reference \ac{mpc}. 
We considered hypothetical prescient load and weather forecasts to rule out the influence of forecast uncertainties. 
Moreover, the reference \ac{mpc} has perfect knowledge of the battery dynamics.
For our model, we use prediction horizon $L = 10$. 
Running the simulation resulted in the system trajectory shown in Figure \ref{fig: NonlinBat Ref trajectory}. The upper figure displays the time series of each unit's power whereas the corresponding trajectory of the stored energy is displayed in the lower figure.

\begin{table}[ht]
\centering
\caption{Parameter values of the islanded grid.} \label{table: Parameter values of simple mpc}
\begin{tabular}{lcclcclcc}
\toprule
 Param. & Value & & Param. & Value & & Param. & Value \\
\cmidrule{1-2}
\cmidrule{4-5}
\cmidrule{7-8}
 $c_{0}$ & 1 & & $p^{max}_{t}$ & 1 & & $B_{s}$  & -0.5 \\
 $c_{1}$& 0.3 & & $p^{min}_{s}$ & -1 & & $B_{q}$ & -0.05 \\
 $c_{2}$& 0.2 & & $p^{max}_{s}$ & 1 & & $x^{min}$ & 0.5 \\
 $\gamma$& 0.9 & & $p^{min}_{r}$ & 0 & & $x^{max}$ & 6.5 \\
 $p^{min}_{t}$& 0.3 & & $A$ & 0.99 & & &\\
 \bottomrule
\end{tabular}
\end{table}

\section{Linear Data-Driven \acs{mpc}} \label{sec: linear mpc}
In this section we control the system from Section \ref{sec: Microgrid Model} with nonlinear battery dynamics using a linear data-driven \ac{mpc} approach.
We consider unknown battery dynamics but assume mild nonlinearities, where the nonlinear terms have a much smaller influence on the dynamics than the linear ones.
This allows us to use a modified version of \eqref{eq: DD Trajectory Criteria} and build a linear data-driven \ac{mpc} that accounts for plant-model-mismatch.

\subsection{Problem Statement}
Let $G$ denote our battery dynamics. 
We assume that $G$ is a controllable \ac{lti} system and that $\tilde{n}$ is an upper approximation  of $G$'s minimal order $n$, i.e., $\tilde{n}\ge n$. 
For the moment, let us assume that we have no plant-model-mismatch, that we can measure the battery's stored energy, i,e., $y(k)=x(k)$, and control its power, $u(k)=p_{s}(k)$. 
Measuring the input and its corresponding output $N$ consecutive times allows us to create the measured sequences $\mathbb{U}^{m}$ and $\mathbb{Y}^{m}$.
If $\mathbb{U}^{m}$ is persistently exciting of order $L+\tilde{n}$, then based on Theorem \ref{theorem: Linear Case}, the \ac{i/o} sequence $\{u(k)\}^{L-1}_{k=-\tilde{n}},\{y(k)\}^{L-1}_{k=-\tilde{n}}$ is a trajectory of $G$ if and only if there exists an $\alpha\in \mathbb{R}^{N-(L+\tilde{n})+1}$ such that

\begin{equation*}
\begin{bmatrix}
\mathbf{U}\\
\mathbf{Y}
\end{bmatrix} = 
\begin{bmatrix}
H_{L+\tilde{n}}(\mathbb{U}^{m})\\
H_{L+\tilde{n}}(\mathbb{Y}^{m})
\end{bmatrix}\alpha
\end{equation*}
with $\mathbf{U}=[u(k)]^{L-1}_{k=-\tilde{n}}$ and $\mathbf{Y}=[y(k)]^{L-1}_{k=-\tilde{n}}$, i.e.,
\begin{equation} \label{eq: MPC linear battery dynamics Xs}
\begin{bmatrix}
[p_{s}(k|t)]^{L-1}_{k=-\tilde{n}}\\ 
[x(k|t)]^{L-1}_{k=-\tilde{n}}
\end{bmatrix} =
\begin{bmatrix}
H_{L+\tilde{n}}(\{p^{m}_{s}(t)\}^{N-1}_{t=0})\\
H_{L+\tilde{n}}(\{x^{m}_{s}(t)\}^{N-1}_{t=0})
\end{bmatrix}\alpha
\end{equation}
for $t > N - 1$.

Recall Remark~\ref{remark: L tilde} and Theorem~\ref{theorem: Linear Case}: 
it is easy to see that if $\mathbb{U}^{m}$ is persistently exciting of order $L+\tilde{n}$, then it is also persistently exciting of order $L+n \le L+\tilde{n}$. 

The first $\tilde{n}$ elements are used to describe the initial conditions of the battery using the equality constraint
\begin{equation} \label{eq: Initial Conditions Linear Case}
\begin{bmatrix}
[p_{s}(k|t)]^{-1}_{k=-\tilde{n}}\\
[x(k|t)]^{-1}_{k=-\tilde{n}}
\end{bmatrix} =
\begin{bmatrix}
[p^{m}_{s}(k)]^{t-1}_{k=t-\tilde{n}}\\
[x^{m}_{s}(k)]^{t-1}_{k=t-\tilde{n}}
\end{bmatrix}. \\
\end{equation}
The remaining $L$ values are used to obtain a forecast for $L$ steps into the future. 

Until now, \eqref{eq: MPC linear battery dynamics Xs} and \eqref{eq: Initial Conditions Linear Case} where formulated assuming that $G$ is \ac{lti}. 
However, $G$ is actually not \ac{lti}.
In order to account for the mismatch between the linear model \ac{i/o} data and the \ac{i/o} data from the nonlinear plant, we employ a modified version of \eqref{eq: MPC linear battery dynamics Xs} in the following \ac{mpc} scheme.
\begin{problem}[Linear data-driven \acs{mpc}] \label{prob:linear}
  \begin{subequations}
    \[
      \min_{P, X, \alpha, \beta} \textstyle\sum\limits_{k=0}^{L-1} \ell(k|t)\cdot\gamma^{k} + c_{\alpha}||\alpha||^{2}_{2} + c_{\beta}||\beta||^{2}_{2}
    \]
    subject to \eqref{eq: wind turbine power output}--\eqref{eq: Storage Power Constraints}, \eqref{eq: power equilibrium constraints} for all $k\in[0,L-1]$, \eqref{eq: Storage Capacity Constraints} for all $k\in[1,L]$ with $\delta(-1|t) = \delta^{m}(t)$  as well as
  \begin{equation} \label{eq: beta equation}
	\begin{bmatrix}
	[p_{s}(k|t)]^{L-1}_{k=-\tilde{n}}\\
	[x(k|t)]^{L-1}_{k=-\tilde{n}} + \beta
	\end{bmatrix} =
	\begin{bmatrix}
	H_{L+\tilde{n}}(\{p^{m}_{s}(t)\}^{N-1}_{t=0})\\
	H_{L+\tilde{n}}(\{x^{m}_{s}(t)\}^{N-1}_{t=0})
	\end{bmatrix}\alpha
	\end{equation}
	\text{for $t > N - 1$ and}
	\begin{equation} \label{eq: initial constraints problem 2}
	\begin{bmatrix}
	[p_{s}(k|t)]^{-1}_{k=-\tilde{n}}\\
	[x(k|t)]^{-1}_{k=-\tilde{n}}
	\end{bmatrix} =
	\begin{bmatrix}
	[p^{m}_{s}(k)]^{t-1}_{k=t-\tilde{n}}\\
	[x^{m}_{s}(k)]^{t-1}_{k=t-\tilde{n}}
	\end{bmatrix}. \\
	\end{equation}
  \end{subequations}
\end{problem}

The measured output sequence $\{x^{m}_{s}(k)\}^{N-1}_{k=0}$ is composed of one part that reflects the linear dynamics and another part which stems from the nonlinear term.
In a way, this nonlinear part can be understood as an error that acts on the linear system.
Therefore, large values of $\alpha$ increase the error that the Hankel matrices with the data from the nonlinear system introduce. 
To overcome this issue, motivated by \cite{Lygeros_InTheShallowsOfDeePC,DDControlWithStabilityGuarantees, Linear_Tracking_MPC_2}, we added the cost term $c_{\alpha}||\alpha||^{2}_{2}$ to punish large values of $\alpha$.
Similar to \cite{Linear_Tracking_MPC_2}, where state-nonlinearities are investigated, we further added the slack variable $\beta\in\mathbb{R}^{p\cdot(L-1+\tilde{n})}$ to \eqref{eq: beta equation} to account for model mismatch and used the term $c_{\beta}||\beta||^{2}_{2}$ to keep $\beta$ small. 
\begin{remark}
	In \cite{DDControlWithStabilityGuarantees}, a similar slack variable is employed to account for noisy measurement data and
	\cite{Lygeros_InTheShallowsOfDeePC,DD_multi_energy_dis_system} added a slack variable to \eqref{eq: initial constraints problem 2} instead of \eqref{eq: beta equation}.
	The latter approach, did however not lead to satisfactory results in our case.	
\end{remark}

\subsection{Case Study} \label{sec: linear case study}
Let us consider the setup from Section \ref{sec: reference case study}.
The input $u(k)=p_{s}(k)$ and output $y(k)=x_{s}(k)$ of the battery were measured throughout the simulation.
We used $185$ consecutive measurements from the simulation to form the measured sequences $\mathbb{U}^{m}=\{u(k)\}^{184}_{k=0}$ and $\mathbb{Y}^{m}=\{y(k)\}^{184}_{k=0}$.
We considered $\tilde{n}=1$ and created the Hankel matrices $H_{L+\tilde{n}}(\mathbb{U}^{m})$ and $H_{L+\tilde{n}}(\mathbb{Y}^{m})$. 
We positively validated persistence of excitation of order $L+\tilde{n}$ for $\mathbb{U}^{m}$. 

We employed Problem~\ref{prob:linear} with $c_{\alpha}=5$ and $c_{\beta}=10^{4}$ in closed-loop simulations.
The average violation of the state of charge constraints of the reference \ac{mpc} is $0$, which was expected, since it has perfect knowledge of the battery's dynamics.
The average violation with Problem \ref{prob:linear} was nonzero but small. 
This illustrates the effectiveness of the cost terms associated with $\alpha$ and slack variable $\beta$.
However, it also shows that the data-driven \ac{mpc} did not manage to perfectly describe the actual input-nonlinear battery dynamics.
Nonetheless, one great advantage of using the data-driven \ac{mpc} is that the dynamics of the system do not need to be explicitly known in the \ac{mpc} formulation and that an identification step can be disregarded.

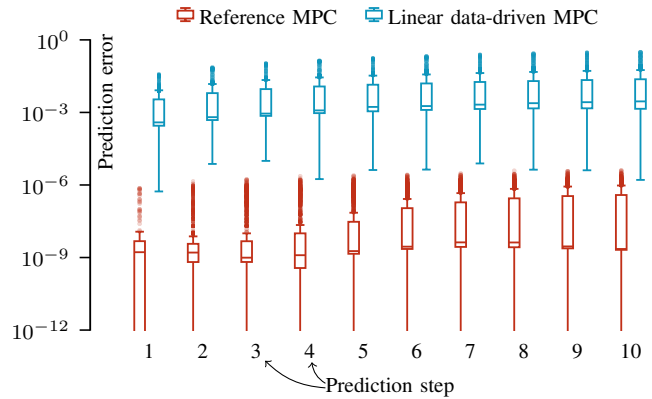
\begin{figure}[ht]
    \centering
    \tikzsetnextfilename{linearDdBoxplot}

\begin{tikzpicture}[font=\footnotesize]
  \begin{semilogyaxis}[
    boxplot semilogy,
    ymax = 1,
    ]

  \addlegendimage{boxplot legend, color = vgRed}
  \addlegendentry{Reference MPC}

  \addlegendimage{boxplot legend, color = vgLightBlue}
  \addlegendentry{Linear data-driven MPC}

  \foreach \step in {0, ..., 9}
  {

    \addplot [linestyle boxplot red, xshift=0.5mm] table[y = Step \step, col sep=comma]
    {simulation-data/box-plot-data-nonlinear-dynamics/box-plots-simple-data.csv};

    \addplot [linestyle boxplot blue,  xshift=-0.5mm] table[y = Step \step, col sep=comma]
    {simulation-data/box-plot-data-nonlinear-dynamics/box-plots-dd-data-linear-aprox.csv};
  }

  \end{semilogyaxis}

  \draw[black, ->, bend left=20] (2.6, -0.76) to (1.8, -0.4);
  \draw[black, ->, bend left=25] (2.6, -0.70) to (2.42, -0.42);

\end{tikzpicture}
    \caption{Box plots of prediction error for reference \ac{mpc} (Problem \ref{prob:reference}) and  linear data-driven \acs{mpc} (Problem \ref{prob:linear}).}
    \label{fig: BoxPLot Battery Linear}
\end{figure}

We also investigated the prediction capabilities.
At each step $t$, each \ac{mpc} outputs the optimal future input sequence $\{p^{*}(k|t)\}^{L-1}_{k=0}$.
Parallel to our actual grid (which is used for closed-loop simulations), we use a second identical plant model to which we apply all $L$ predicted optimal values at each step.
This allows to compare the \ac{mpc}s' calculated trajectories $[x^{*}(k|t)]^{L}_{k=1}$ with a perfectly accurate model and calculate the prediction error of the respective \ac{mpc} for each step.
For the simulations performed in Sections~\ref{sec: reference mpc} and~\ref{sec: linear mpc}, the box plots of the errors can be found in Figure~\ref{fig: BoxPLot Battery Linear}. 
Despite the superiority of the reference \ac{mpc}, the data-driven \ac{mpc} still manages to do a decent job in predicting the battery's state trajectory.
However, the prediction capabilities of data-driven \ac{mpc} can be further improved, as illustrated in the next section.

\section{Hammerstein-type data-driven \ac{mpc}} \label{sec: Hammerstein type data-driven MPC}
In this section, we formulate a data-driven \ac{mpc} where only the nature of the battery's nonlinearity is known.
First, we present the \ac{mpc} problem. 
Then, we use it in closed-loop simulations. 

\subsection{Problem Statement}

System \eqref{func: Storage Dynamics Quadratic Default} is a Hammerstein system (see Section \ref{sec: hammerstein}) with $B=1$, $\psi(u)=\alpha_{1}u + \alpha_{2}u^{2}$, $\alpha_{1}=B_{l}$ and $\alpha_{2}=B_{q}$.
To model this system in a data-driven fashion we define the auxiliary linear input vector.

\begin{equation} \label{eq: linear-input example}
    v(k) = \begin{bmatrix}
    \psi_{1}(u(k)) \\
    \psi_{2}(u(k)) \\
    \end{bmatrix} = \begin{bmatrix}
    u(k) \\
    u(k)^{2} \\
    \end{bmatrix}.
\end{equation}

If the nature of the nonlinearity is known, then (based on Theorem~\ref{theorem: Nonlinear Case}) we can express the system behavior by 
\begin{equation} \label{eq: Nonlinear Dynamics Constraint}
    \begin{bmatrix}
    \mathbf{V}\\
    \mathbf{Y}
    \end{bmatrix}=\begin{bmatrix}
    H_{L+\tilde{n}}(\mathbb{V}^{m})\\
    H_{L+\tilde{n}}(\mathbb{Y}^{m})
    \end{bmatrix}\alpha
\end{equation}
with
$\mathbf{V}=[v(k|t)]^{L-1}_{k=-\tilde{n}}$,
$\mathbf{Y}=[y(k|t)]^{L-1}_{k=-\tilde{n}}$,
$\mathbb{V}^{m}=\{v^{m}(t)\}^{N-1}_{t=0}$,
$v^{m}(t)=[\psi_{1}(u^{m}(t))~\psi_{2}(u^{m}(t))]^{T}$,
$\mathbb{Y}^{m}=\{y^{m}(t)\}^{N-1}_{t=0}$,
for $t  > N - 1$ and the initial conditions by
\begin{equation} \label{eq: Nonlinear Initial Conditions Constraint}
\begin{bmatrix}
[v(k|t)]^{-1}_{k=-\tilde{n}}\\
[y(k|t)]^{-1}_{k=-\tilde{n}}
\end{bmatrix}=\begin{bmatrix}
[v^{m}(k)]^{t-1}_{k=t-\tilde{n}}\\
[y^{m}(k)]^{t-1}_{k=t-\tilde{n}}
\end{bmatrix}.
\end{equation}

For this approach to work, one has also to consider the relationship between the two basis functions. 
In this example, this means that we need to enforce the equality constraint 
\begin{equation} \label{eq: nonlinear equality constraint}
    \psi_{2}(u(k|t)) = \psi_{1}(u(k|t))^{2}
\end{equation}
for all $k\in \begin{bmatrix}
-\tilde{n}, L-1
\end{bmatrix}$.
If $v(k|t)=[v_{1}(k|t)~v_{2}(k|t)]^{T}$, then another way of formulating (\ref{eq: nonlinear equality constraint}) is by
\begin{equation} \label{eq: linear-input constraint}
    v_{2}(k|t) = v_{1}(k|t)^{2}.
\end{equation}

Fortunately, available commercial solvers, e.g., Gurobi, can handle such quadratic constraints.
In what follows, let 
$\tilde{p}(k|t)=\begin{bmatrix}p_{t}(k|t)~v_{1}(k|t)~v_{2}(k|t)~p_{r}(k|t)\end{bmatrix}^{T}$
and
$\tilde{P}=[\tilde{p}(k|t)]^{L-1}_{k=-\tilde{n}}$.
We can then formulate the following \ac{mpc} problem.
\begin{problem}[Hammerstein-type data-driven \acs{mpc}] \label{prob:nonlinear}
  \begin{subequations}
    \[
      \min_{\tilde{P}, X, \alpha} \textstyle\sum\limits_{k=0}^{L-1} \ell(k|t)\cdot\gamma^{k}
    \]
    subject to \eqref{eq: wind turbine power output}--\eqref{eq: Storage Power Constraints}, \eqref{eq: power equilibrium constraints} for all $k\in[0,L-1]$, \eqref{eq: Storage Capacity Constraints} for all $k\in[1,L]$, and the constraints
    \begin{equation}
      p_{t}(k|t) + v_{1}(k|t) + p_{r}(k|t) + w_{d}(k|t) = 0,
  \end{equation} 
    \begin{equation}
      v_{2}(k|t) = v_{1}(k|t)^{2}, 
    \end{equation}
    for all $k\in[-\tilde{n},L-1]$
    with $\delta(-1|t)=\delta^{m}(t)$, as well as
  \begin{equation}
    \begin{bmatrix}
    [v(k|t)]^{L-1}_{k=-\tilde{n}}\\
    [y(k|t)]^{L-1}_{k=-\tilde{n}}
    \end{bmatrix}=\begin{bmatrix}
    H_{L+\tilde{n}}(\mathbb{V}^{m})\\
    H_{L+\tilde{n}}(\mathbb{Y}^{m})
    \end{bmatrix}\alpha
	\end{equation}

	\begin{equation}
		\begin{bmatrix}
			[v(k|t)]^{-1}_{k=-\tilde{n}}\\
			[y(k|t)]^{-1}_{k=-\tilde{n}}
		\end{bmatrix}=\begin{bmatrix}
			[v^{m}(k)]^{t-1}_{k=t-\tilde{n}}\\
			[y^{m}(k)]^{t-1}_{k=t-\tilde{n}}
		\end{bmatrix}.
	\end{equation}
  \end{subequations}
\end{problem}
 
\subsection{Case Study}
\begin{figure}[ht]
    \centering
    \tikzsetnextfilename{nonlinearDdBoxplot}

\begin{tikzpicture}[font=\footnotesize]
  \begin{semilogyaxis}[
    boxplot semilogy,
    ymax = 1,
    ]

  \addlegendimage{boxplot legend, color = vgRed}
  \addlegendentry{Reference MPC}

  \addlegendimage{boxplot legend, color = vgLightBlue}
  \addlegendentry{Hammerstein-type data-driven MPC}

  \foreach \step in {0, ..., 9}
  {

    \addplot [linestyle boxplot red, xshift=0.5mm] table[y = Step \step, col sep=comma]
    {simulation-data/box-plot-data-nonlinear-dynamics/box-plots-simple-data.csv};

    \addplot [linestyle boxplot blue,  xshift=-0.5mm] table[y = Step \step, col sep=comma]
    {simulation-data/box-plot-data-nonlinear-dynamics/box-plots-dd-data.csv};

  }

  \end{semilogyaxis}

  \draw[black, ->, bend left=20] (2.6, -0.76) to (1.8, -0.4);
  \draw[black, ->, bend left=25] (2.6, -0.70) to (2.42, -0.42);

\end{tikzpicture}
    \caption{Box plots of prediction error for reference \ac{mpc} and Hammerstein-type data-driven \acs{mpc}.}
    \label{fig: BoxPLot Battery Nonlinear}
\end{figure}
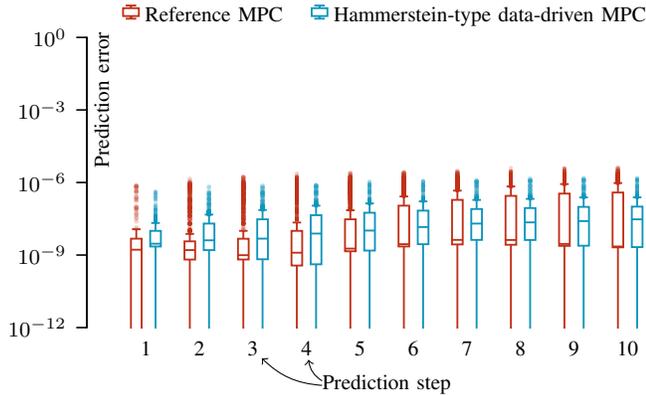

Analogously to Section \ref{sec: linear case study}, we simulated the grid and used  $N=185$ \ac{i/o} measurements to create the data-driven \ac{mpc}. 
Results indicate that the controller based on Problem \ref{prob:nonlinear} yields identical performance as the one based on Problem \ref{prob:reference}.
Moreover, constraints are not violated.
The absolute prediction error of both controllers is below the numerical precision of the solver (see Figure \ref{fig: BoxPLot Battery Nonlinear}) which highlights the ability of Problem \ref{prob:nonlinear} to accurately predict the future state trajectory of a Hammerstein system given that the basis functions in (\ref{eq: linear-input example}) are known.
Problem \ref{prob:reference} had perfect knowledge of the plant's dynamics which most of the time prerequisites a system identification step.
This is not required in Problem \ref{prob:nonlinear}, where the measured data is directly used inside the control loop.
Furthermore, the control scheme of Problem \ref{prob:nonlinear} can learn changing battery dynamics (e.g. $A$ and $B_{q}$ may change due to degradation) by updating its Hankel matrices using more recent measurements. 

\section{Conclusions} \label{sec: conclusions}
This paper discussed how novel data-driven \ac{mpc} schemes can be employed to control power grids with nonlinear Hammerstein-type storage dynamics. 
First, the unknown nonlinearity of the battery was considered to be mild, yet the data-driven \ac{mpc} managed to control the nonlinear plant with small constraint violations.  
Subsequently, the structure of the nonlinearity was considered to be known.
Using an extended version of Willems' fundamental lemma, we developed a data-driven \ac{mpc} that successfully controlled the battery with identical results as the reference \ac{mpc} with perfect knowledge of the dynamics.  
Both data-driven approaches allowed to skip a system identification step and to directly formulate \ac{mpc} schemes from \ac{i/o} data.
 
Future work will focus on theoretical considerations for the case of "mild" input nonlinearities and on tests with more realistic \ac{i/o} data of real-world batteries.

\bibliographystyle{IEEEtran}
\bibliography{references}

\end{document}